# Aid to percutaneous renal access by virtual projection of the ultrasound puncture tract onto fluoroscopic images

Mozer P., Conort P., Leroy A., Baumann M., Payan Y., Troccaz J., Chartier-Kastler E. & Richard F.

# Introduction

Percutaneous renal access in the context of percutaneous nephrolithotomy (PCNL) is a difficult technique, requiring rapid and precise access to a particular calyx. Optimization of the puncture tract by targeting a renal papilla has been demonstrated to decrease the risk of bleeding[1].

At the present time, in routine clinical practice, percutaneous renal access is usually performed with fluoroscopic and/or ultrasound guidance. Each of these imaging modalities presents a number of disadvantages. Fluoroscopy provides images in various planes, but only in 2 dimensions, and exposes the patient and the medical team to x-rays, while ultrasound often provides a blurred image of the kidney and the target, but with no risks for the patient or staff.

In order to minimize these disadvantages, some teams use CT for difficult cases[2], but percutaneous renal access with this tool, apart from being time-consuming, remains difficult as kidney movements cannot be taken into account in real time.

To our knowledge, only two tools are currently under development as an aid to percutaneous renal access. The first system, called PAKY[3], is a joystick-controlled robotic arm. During a ventilatory pause, the operator directs the robot to a target identified on fluoroscopic images. Studies are currently underway to automate CT-guided puncture. The second system is an augmented reality tool[4] that projects the preoperative CT scan onto the patient's body. This system does not allow real-time monitoring of the procedure.

We believe it is possible to improve the precision of percutaneous renal access by using computerized methods to virtually project the ultrasound puncture tract onto previously acquired fluoroscopic images. The procedure can then be performed by a human operator guided by the computer screen or by a robot.

# Objective

The objective of this study was to establish a real-time correspondence between the ultrasound puncture tract and fluoroscopic images acquired at the beginning of the procedure. Such a system would allows better planning of the puncture tract, guided simultaneously by two imaging modalities. The ultrasound puncture tract projected onto fluoroscopic images is only virtual and is visualized when the fluoroscope is no longer is in the operative field.

We present the results of our laboratory experiments on a phantom designed to validate this approach, together with the preliminary results obtained in one patient.

# Material and method

In order to dilate the pyelocaliceal cavities (PCC) and make them visible on fluoroscopy, we initially place a ureteric catheter in the upper calyx according to the usual PCNL procedure. The patient is then placed in the prone position and the PCC are filled with contrast agent mixed with blue-coloured indigo carmine.

The principle of image fusion is based on the possibility of spatially localizing the fluoroscope and the ultrasound transducer in the operating room (Figure 1). We use a Polaris® infrared camera system (Northern Digital Inc, Ontario, Canada), which visualizes markers placed on the various instruments. This type of localizer is already used routinely in the operating room by surgical navigation systems, especially for orthopaedic surgery. It provides the spatial position of the surgical instruments in real time with a precision of the order of one millimetre.

We attach a calibration target onto the fluoroscope and the rigid-body ultrasound transducer (Figure 2, left). A so-called reference rigid-body must be fixed to the operating table to determine the relative positions of the fluoroscope and the ultrasound transducer throughout the operation. The Polaris® system therefore indicates the real-time position of the ultrasound transducer with respect to the fluoroscopic image. As we want to visualize the ultrasound puncture tract on fluoroscopic images, preliminary calibration of this tract must be performed. This calibration step, which takes less than a minute, consists of placing a wire in the needle guide and determining its course by presenting a so-called calibrator to the Polaris® (Figure 2, right). We use an ultrasound needle guide allowing the choice between three different nephrostomy tracts.

The first step consists of acquisition of fluoroscopic images. As the kidney moves with breathing, the anaesthetist is asked to programme the ventilator to stop ventilation for less than 10 seconds at the end of expiration so that we can perform the first image acquisition. Images can be obtained in several views (-30° and + 30°) by simply asking the anaesthetist to programme another end-expiratory ventilatory pause. Fluoroscopic images acquired in several views are particularly useful to visualize all of the calices and to determine the axis of caliceal stalks. The fluoroscope is then removed from the operative field to leave more room for the operator.

From this point on, the computer system which has integrated the spatial position of fluoroscopic images is able to represent the nephrostomy tract selected on the ultrasound transducer. Only the ultrasound transducer, fitted with its puncture guide, is then mobilized. The operator therefore visualizes, in real time, both the puncture tract on the ultrasound image and its virtual tract on all fluoroscopic views. The kidney obviously moves on the ultrasound image, but the anaesthetist can always be asked to perform a ventilatory pause with the same insufflated volume as during fluoroscopic acquisition to ensure perfect image fusion. The operator is therefore able to determine the optimal pucture tract and can visualize progression of the needle on the ultrasound monitor. The operator consequently utilizes the respective performances of each imaging system simultaneously with minimal x-ray exposure. These two imaging systems therefore provide a real synergy, while leaving the surgeon completely free to perform the operative procedure.

## Results

We initially validated our approach by using a phantom manufactured in the laboratory (Figure 3). This phantom is composed of agar gel containing 2 metal objects acting as targets which can be visualized by both ultrasound and fluoroscopy. We verified, by puncturing these targets, that they are superimposed on both fluoroscopy and ultrasound. As the phantom is immersed in water and as the agar gel of the phantom is transparent, we were also able to verify the "real" position of the needle tip in the metal objects. We estimated visually that the correspondence between the two imaging modalities was of the order of one millimetre (Figure 4).

After approval by our Ethics Committee, we used this system to access the pyelocaliceal cavities in one patient.

Figure 5 represents the correspondence of the puncture tract on ultrasound and fluoroscopic images. The operator, with considerable experience in ultrasound, verified that the calices visible on the ultrasound puncture tract actually corresponded to the calices visible on the virtual fluoroscopic puncture tract. A single puncture was necessary to reach the target and the needle tract visualized by a fluoroscopic image of the needle corresponded to that decided by the operator. No intraoperative difficulty or complication was observed.

## Discussion

This idea of superimposing the ultrasound puncture tract onto fluoroscopic images is derived from preliminary studies on ultrasound and CT image fusion performed in our laboratory[5].

We have demonstrated the possibility of fusing images of the kidney derived from these two imaging modalities by segmenting the renal capsule on CT and ultrasound images. However, this is impossible in routine clinical practice, as, although segmentation can be automated for CT images, this is much more difficult for ultrasound images, on which the kidney contours must be segmented manually on several ultrasound images, which would appear difficult to perform routinely in the operating room.

Our solution of fusion of easily acquired images that can be performed in any operating room must be robust, as simple superimposition of the ultrasound puncture tract overcomes the problems of correspondence between CT and ultrasound images.

Kidney motion can be considered to follow respiratory motion. It is very easy for the operator to determine a precise moment of the respiratory cycle to perform puncture, for example at the end of expiration. Once the needle has entered the renal capsule and parenchyma, the kidney remains virtually immobile, thereby limiting differences between the two imaging sources. In practice, the anaesthetist must stop the ventilator at the same insufflated volume as during acquisition of fluoroscopic images.

## Conclusion

This system can therefore be used to virtually project the ultrasound nephrostomy tract onto fluoroscopic images. This system is obviously generic and is not limited to percutaneous renal access, but could also be used for all applications in which the target is visible on both imaging modalities.

The result obtained on the first patient has encouraged us to initiate a prospective study on the value of this system in routine clinical practice.

Figures

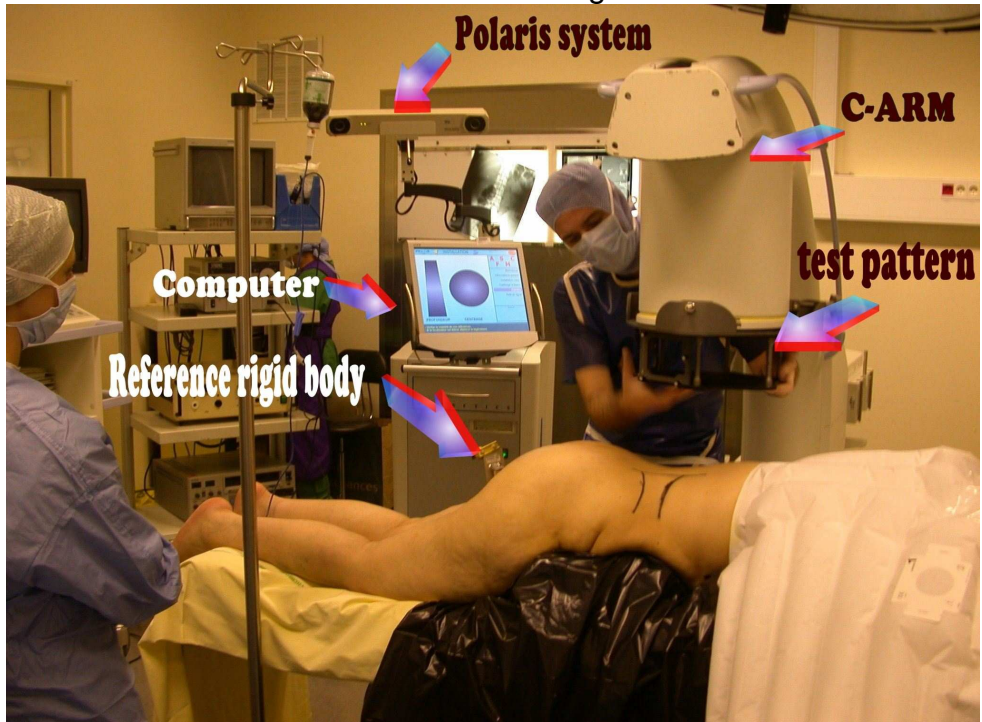

Figure 1: Operative set-up

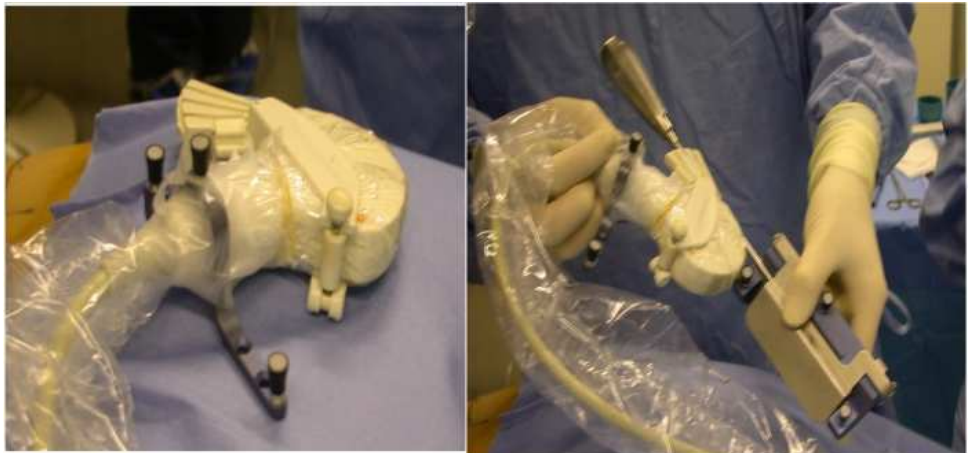

Figure 2: Left: rigid-body ultrasound transducer. Right: Calibration of the nephrostomy tract.

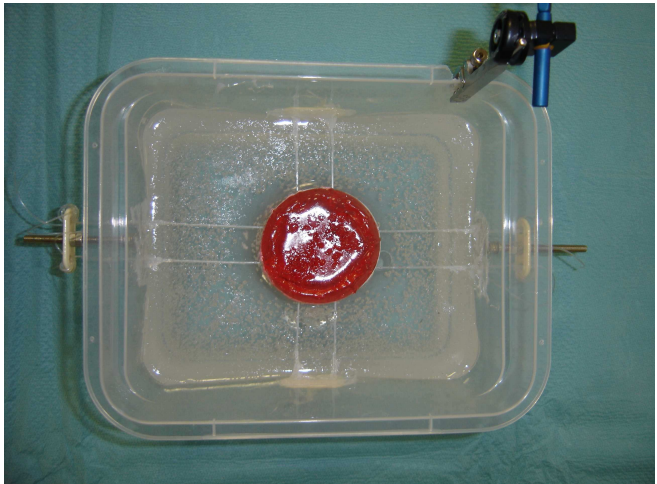
Figure 3: Phantom.

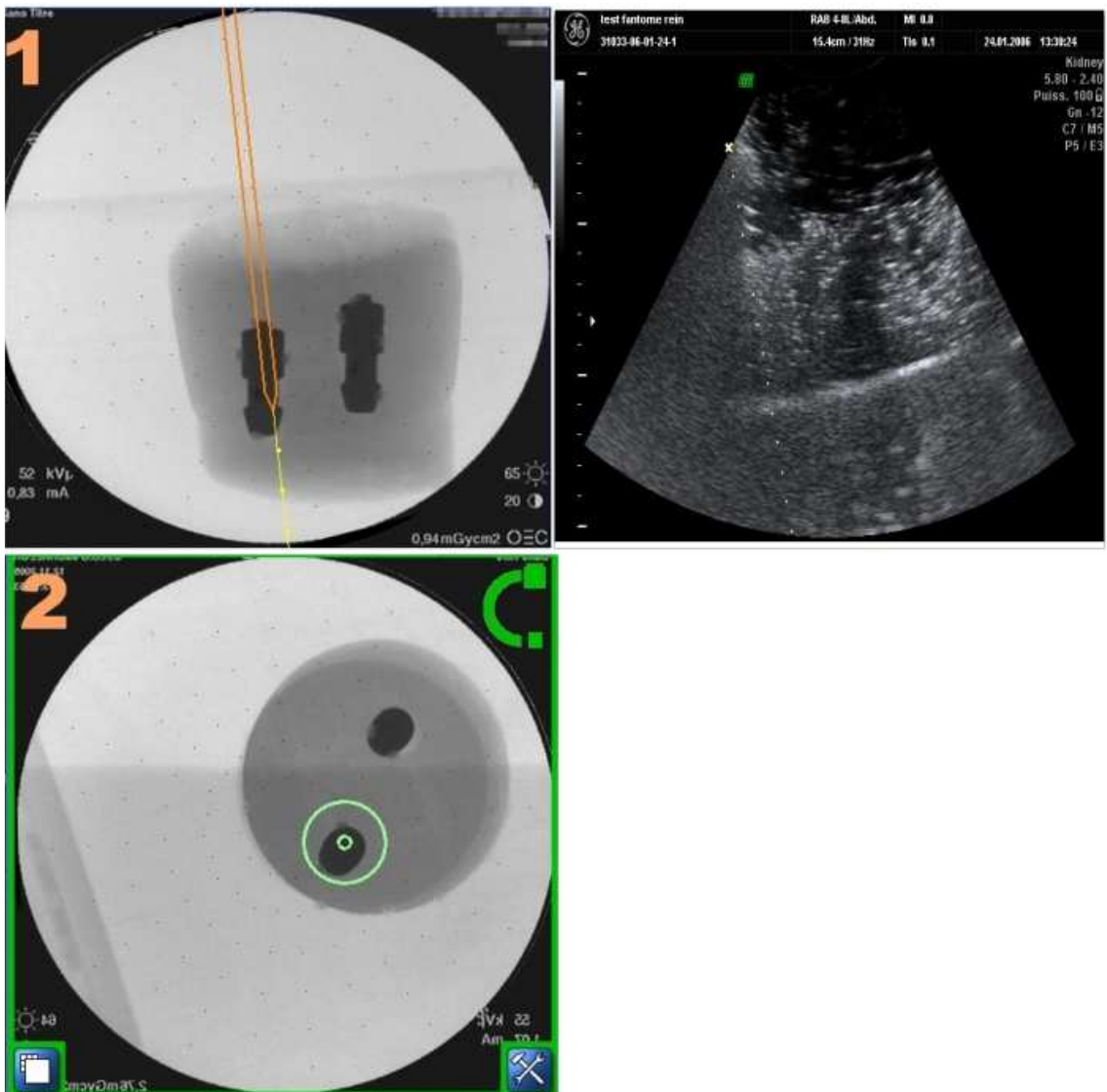
Figure 4: Phantom test. Left: fluoroscopic images. Right: ultrasound image.

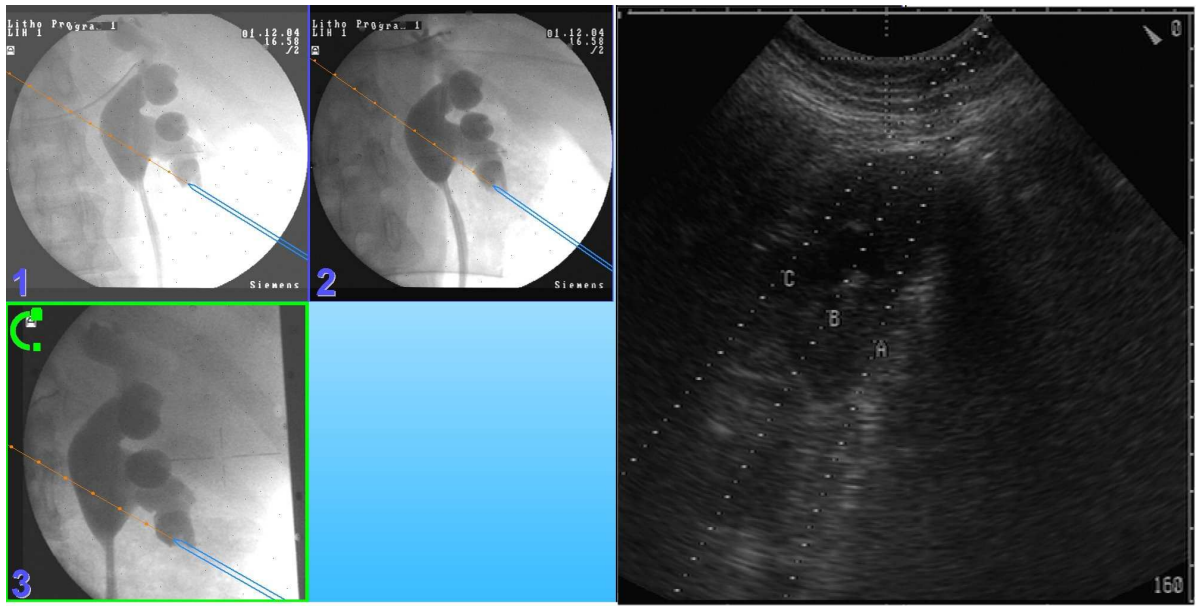
Figure 5: Test on patient. Left: fluoroscopic images. Right: ultrasound image.